\documentclass{PoS}
\pdfoutput=1

\title{Recent Results from BESIII}

\ShortTitle{Recent Results from BESIII}

\author{\speaker{Liaoyuan DONG}
        \thanks{on behalf of the BESIII Collaboration}\\
        Institute of High Energy Physics\\Beijing 100049, China\\
        E-mail: \email{dongly@ihep.ac.cn}}

\abstract{In this talk, we will present some recent results on charmonium spectroscopy and hadron spectroscopy from BESIII experiment,
including the measurements of the masses and widths of $h_c$, $\eta_c$, $\eta_c(2S)$ and some new resonances around 2 GeV.
The results are based on a data sample of 106 million $\psi^\prime$ events and 226 million J/$\psi$ events
collected with the BESIII detector at the BEPCII collider.
}

\FullConference{The 2011 Europhysics Conference on High Energy Physics-HEP 2011,\\
		July 21-27, 2011\\
		Grenoble, Rhône-Alpes France}

\begin{document}

\section{Introduction}
In this paper, some recent results from BESIII experiment on charmonium spectroscopy and hadron spectroscopy
are presented based on about 106 million $\psi^\prime$ events and 226 million J/$\psi$ events
collected by the BESIII detector at the BEPCII collider.

\section{Observation of $h_c$}
\begin{figure}[htb]
  \begin{center}
    \includegraphics[width=0.4\textwidth]{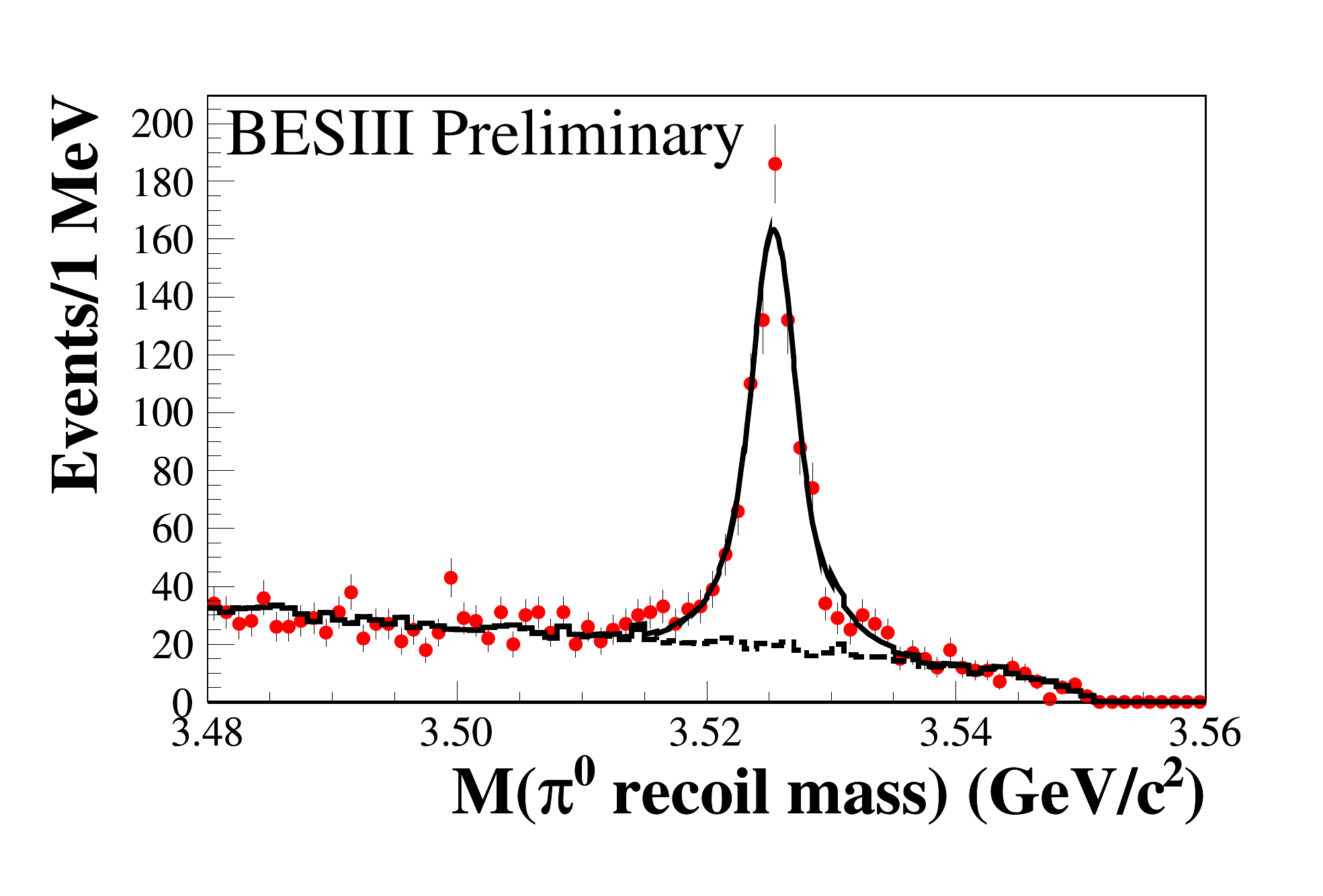}
    \caption{The summed $\pi^0$ recoil-mass spectrum of 16 specific decay processes of $\eta_c$ in the decay chain of $\psi^\prime
             \rightarrow \pi^0 h_c$, $h_c \rightarrow \gamma \eta_c$, where the line is the fit result.}
    \label{fig:hc_sum}
  \end{center}
\end{figure}

Using the largest $\psi'$ sample collected by the BESIII,
we study the 16 specific decay processes of $\eta_c$ in the decay chain of $\psi^\prime
\rightarrow \pi^0 h_c$, $h_c \rightarrow \gamma \eta_c$.
Figure~\ref{fig:hc_sum} shows the $\pi^0$ recoil-mass spectrum of the sum of the 16 decay modes.
A simultaneous fit to the 16 $\pi^0$ recoil-mass spectra
yields $M(h_c)=3525.31\pm0.11\pm0.15\mathrm{MeV/c^2}$ and
$\Gamma(h_c)=0.70\pm0.28\pm0.25\mathrm{MeV}$.
These preliminary results are consistent with the previous BESIII inclusive measurement~\cite{hc-inc-bes3}.

\section{Measurement of the $\eta_c$ properties}
\begin{figure}[htb]
  \begin{center}
    \includegraphics[width=0.85\textwidth]{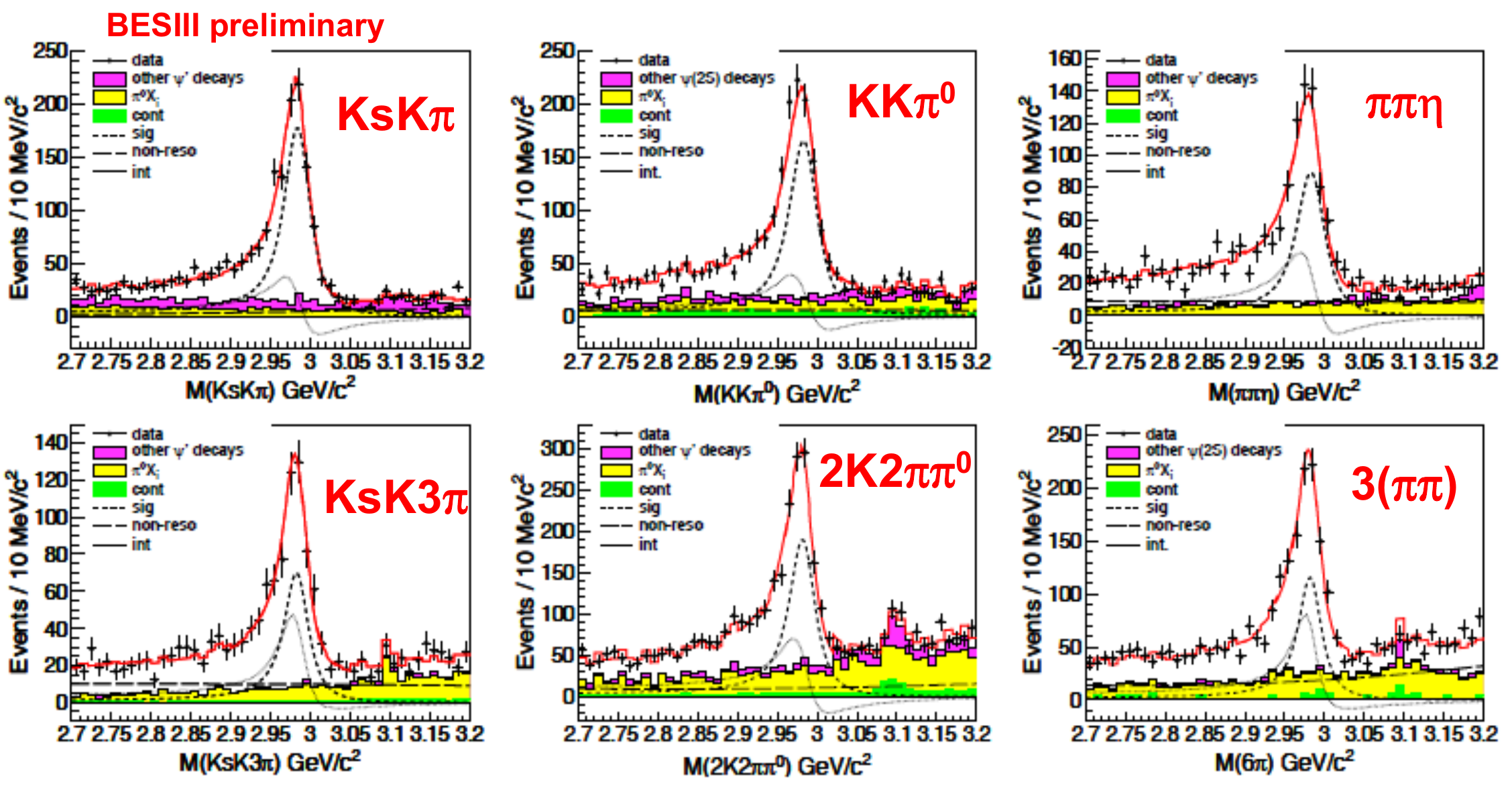}
    \caption{The mass spectra for different decay modes, where the line is the result of the simultaneous fit.}
    \label{fig:etac_fit}
  \end{center}
\end{figure}

Based on the $\psi'$ sample, the $\eta_c$
mass and width are measured in the radiative transition
$\psi'\to\gamma\eta_c$, where $\eta_c$ are reconstructed from 
six decay modes:
$K_S^0K\pi$, $K^+K^-\pi^0$, $\pi^+\pi^-\eta$,
$K_S^0K3\pi$, $K^+K^-\pi^+\pi^-\pi^0$ and $3(\pi^+\pi^-)$.
A simultaneous fit with the unique $\eta_c$ mass and width is
performed on the $\eta_c$ mass spectra, where the interference
between $\eta_c$ and non-resonance decays is considered and the
quantum number of the non-$\eta_c$ components are assumed to be
$0^{-+}$. Assuming an universal relative phase between the two
amplitudes, we obtain $\eta_c$ mass and width, $M = 2984.2 \pm 0.6
\pm 0.5$ MeV/$c^2$ and $\Gamma = 31.4 \pm 1.2 \pm 0.6$ MeV,
respectively, as well as the relative phase $\phi = 2.41 \pm
0.06\pm0.04$ rad. Figure~\ref{fig:etac_fit} shows the fit results
in the six $\eta_c$ decay modes.

\section{Observation of the M1 transition $\psi'\to\gamma\eta_c(2S)$}

\begin{figure}[htb]
  \begin{center}
    \includegraphics[width=0.40\textwidth]{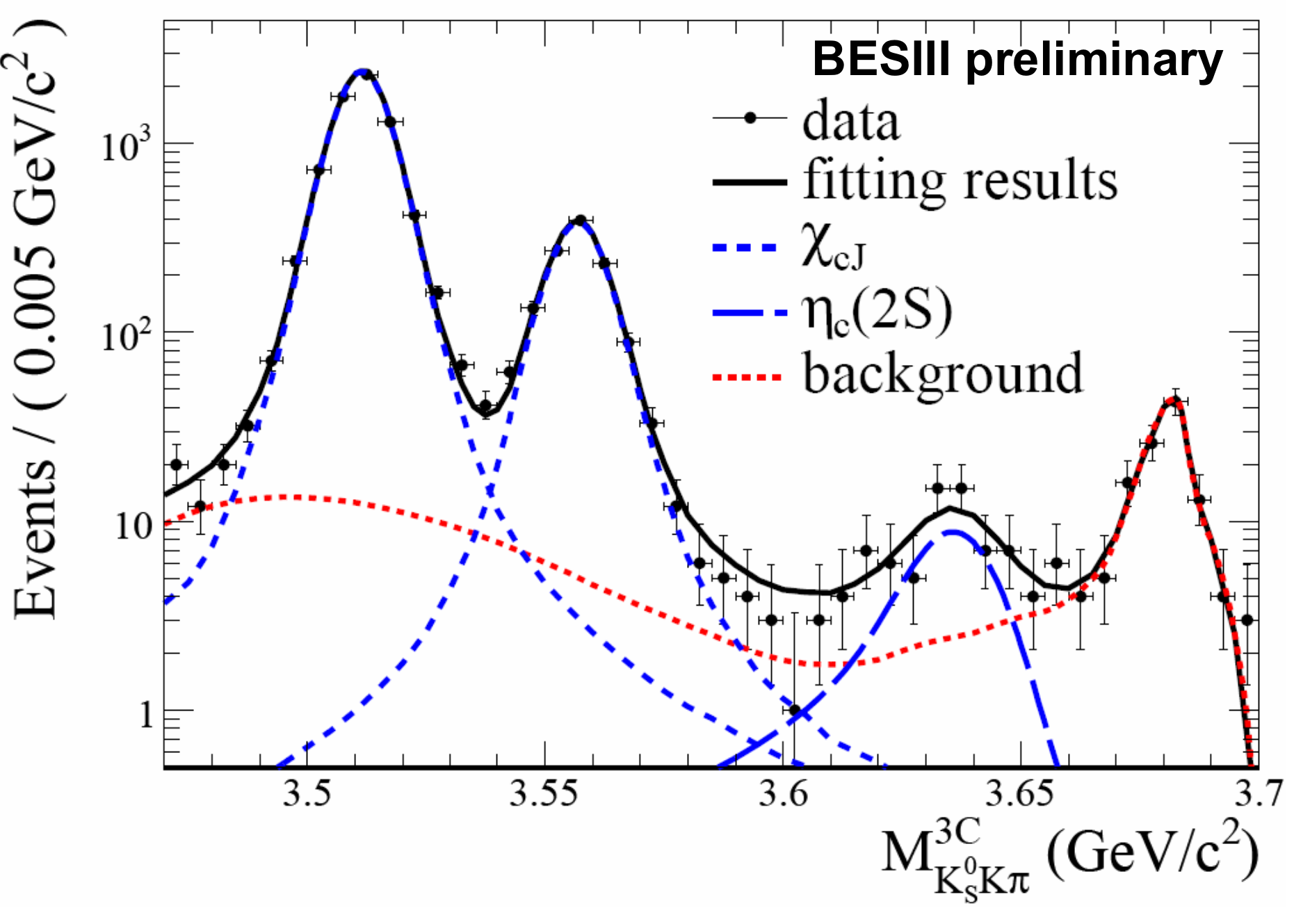}
    \caption{The invariant mass spectrum of $K_S^0K\pi$ from $\psi'\to\gamma K_S^0K\pi$.}
    \label{fig:etacp}
  \end{center}
\end{figure}

BESIII observed the M1 transition $\psi'\to\gamma\eta_c(2S)$ with
the decay mode $\eta_c(2S)\to K_S^0K\pi$.
Figure~\ref{fig:etacp} shows the preliminary result for the
$K_S^0K\pi$ invariant mass distribution, here the
three-constraints kinematic fit has been applied (the energy
of the photon is allowed to be floating). 
With the width of $\eta_c(2S)$ fixed to PDG value,
we  measure mass of
$\eta_c(2S)$ to be $3638.5 \pm 2.3 \pm 1.0$ MeV/$c^2$ and
the $BR(\psi^\prime \rightarrow \gamma
\eta_c(2S))\times BR(\eta_c(2S) \rightarrow K_sK^\pm\pi^\mp)$ to be $(2.98
\pm 0.57\pm 0.48) \times 10^{-6}$. 
The statistical significance for the M1 transition
$\psi^\prime \rightarrow \gamma \eta_c(2S)$ is more than 6.0$\sigma$.
Combining the result $B(\eta_c(2S)\to K\overline{K}\pi)=(1.9\pm0.4\pm1.1)\%$ from BABAR~\cite{etacp-babar},
the M1 transition rate for
$\psi^\prime \rightarrow \gamma \eta_c(2S)$ is derived as
$BR(\psi^\prime \rightarrow \gamma \eta_c(2S))= (4.7 \pm 0.9 \pm
3.0)\times 10^{-4}$, which   
is consistent with the CLEOc's upper limit~\cite{etacp-cleo-c}.

\section{Observation of new resonances in $J/\psi \rightarrow \gamma \eta^\prime \pi^+\pi^-$ and in $J/\psi \rightarrow \omega
\eta\pi^+\pi^-$}

\begin{figure}[htb]
  \begin{center}
    \includegraphics[width=0.7\textwidth]{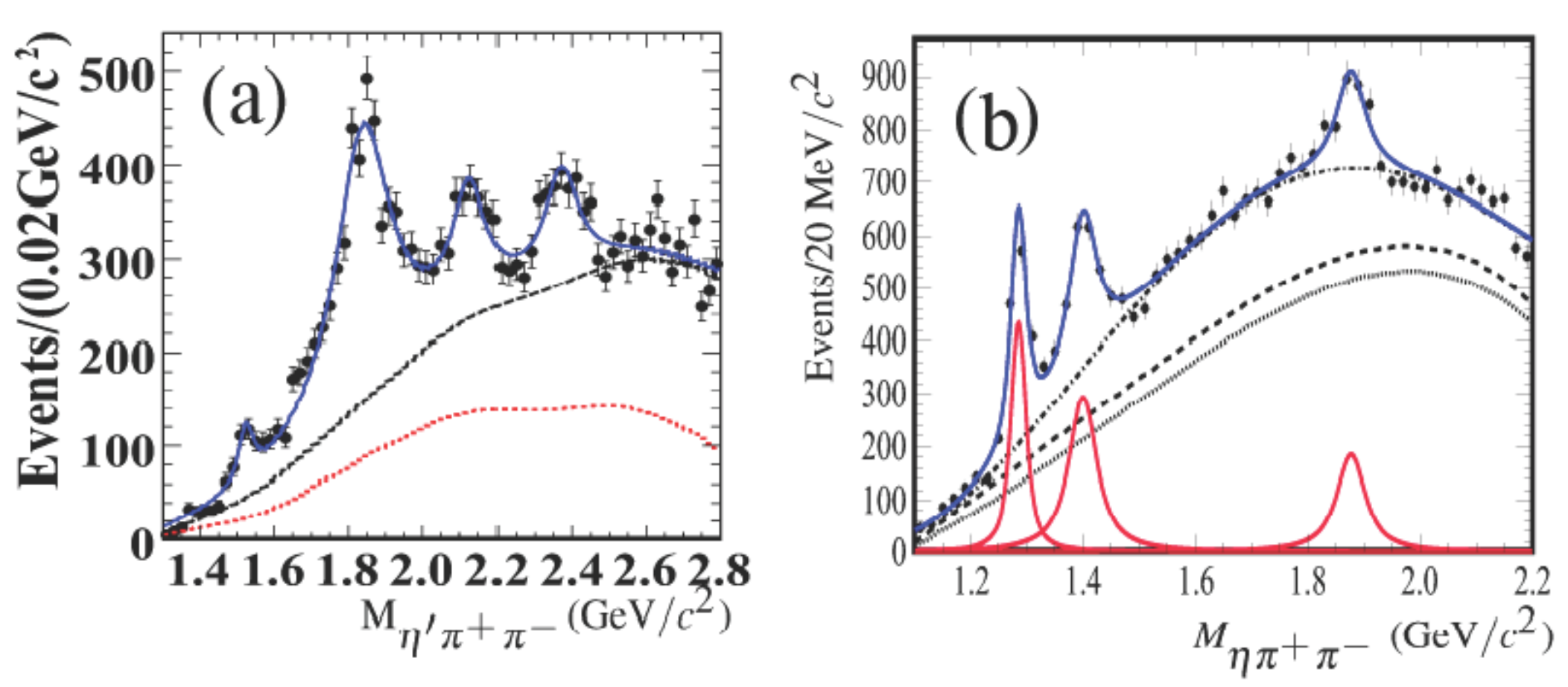}
    \caption{(a) The $\eta^\prime\pi^+\pi^-$ invariant mass
    distribution for the selected $J/\psi\to\gamma\eta^\prime\pi^+\pi^-$ events from the two $\eta^\prime$ decay modes,
    (b) the $\eta \pi^+\pi^-$ invariant mass    
    distribution for the selected $J/\psi\to\omega\eta \pi^+\pi^-$ events in $a_0(980)$ mass window.
    Figures are taken from Refs.~\cite{bes3-x1835} and ~\cite{bes3-x1870}, and described there in more detail.}
    \label{fig:x18xx}
  \end{center}
\end{figure}

The X(1835) was first observed in the $J/\psi \rightarrow \gamma \eta^\prime \pi^+\pi^-$ with
statistical significance of $7.7\sigma$ by the BESII experiment.
A high statistical $J/\psi$ data sample collected with the BESIII
provides an opportunity to
confirm the existence of the X(1835)
and search for other new resonances.
The $\eta^\prime \pi^+\pi^-$ invariant mass
spectrum for the combined two $\eta^\prime$ decays, $\eta^\prime \rightarrow \gamma \rho$ and $\eta^\prime \rightarrow 
\pi^+\pi^- \eta$, is presented in Figure~\ref{fig:x18xx}(a).
The $X(1835)$ resonance is clearly seen. Additional peaks are observed around 2.1 and 2.3  
GeV/c$^2$, denoted as $X(2120)$ and $X(2370)$.
The mass and width of X(1835) are measured to be $M=1836.5 \pm 3.0^{+5.6}_{-2.1}$
$\mathrm{MeV/c^2}$ and $\Gamma = 190 \pm 9^{+38}_{-36}\mathrm{MeV}$ with
significance larger than 20$\sigma$. The mass and width for  
$X(2120)$ ($X(2370)$) are determined to be $M= 2122.4 \pm
6.7^{+4.7}_{-2.7}\mathrm{MeV/c^2}$ ($M=2376.3\pm 8.7^{+3.2}_{-4.3}
\mathrm{MeV/c^2}$) and $\Gamma = 83\pm 16^{+31}_{-11}\mathrm{MeV}$ ($\Gamma = 83\pm
17^{+44}_{-6} \mathrm{MeV}$) with significance of 7.2$\sigma$(6.4$\sigma$). 
For more details, we refer to Ref.~\cite{bes3-x1835}.

The decay J/$\psi \rightarrow \omega \eta \pi^+\pi^-$, in which the 
$\omega$ decays to $\pi^+\pi^-\pi^0$ and the $\eta /\pi^0$ decays to
a pair of photons, is studied to search for the
X(1835). The  $\eta \pi^+\pi^-$ invariant mass spectrum with events in the     
$a_0(980)$ mass window is shown in Figure~\ref{fig:x18xx} (b). Both   
$f_1(1285)$ and $\eta(1405)$ are observed significantly. A clear    
peak around 1900 MeV, denoted as $X(1870)$ is also seen.
A fit with three resonances with simple BW formula
yields a mass $M=1877.3 \pm 6.3^{+3.4}_{-7.4}\mathrm{MeV/c^2}$ and a width
$\Gamma = 57\pm 12^{+19}_{-4}\mathrm{MeV}$ for the $X(1870)$ structure with 
statistical significance of 7.2$\sigma$. 
More details on the data analysis can be found in Ref.~\cite{bes3-x1870}.

\section{Summary and outlook}

Some recent results on charmonium spectroscopy and hadron spectroscopy from BESIII experiment are shown
based on a data sample of 106 million $\psi^\prime$ and about 226 million $J/\psi$ events.
In 2010 and 2011, BESIII have acquired nearly 3 $\ensuremath{\mbox{fb}^{-1}}$ of data at the $\psi(3770)$ resonance.
This sample allows BESIII to begin the charm physics program, including the
mixing and CP violation studies, as well measurements of absolute branching fractions and studies of semi-leptonic decays.


\begin{thebibliography}{99}

\bibitem{hc-inc-bes3}  M. Ablikim         {\it et al.} (BESIII Collaboration), {\it Phys. Rev. Lett.} {\bf 104}, 132002 (2010).
\bibitem{etacp-babar}  B. Aubert          {\it et al.} (BABAR Collaboration),  {\it Phys. Rev. D} {\bf 78}, 012006 (2008).
\bibitem{etacp-cleo-c} D. Cronin-Hennessy {\it et al.} (CLEOc Collaboration),  {\it Phys. Rev. D} {\bf 81}, 052002 (2010).
\bibitem{bes3-x1835}   M. Ablikim         {\it et al.} (BESIII Collaboration), {\it Phys. Rev. Lett.} {\bf 106}, 072002 (2011).
\bibitem{bes3-x1870}   M. Ablikim         {\it et al.} (BESIII Collaboration), {\it Phys. Rev. Lett.} {\bf 107}, 182001 (2011).

\end{thebibliography}
\end{document}